\documentclass[aps,prb,twocolumn,showpacs]{revtex4}
\usepackage{graphicx}
\begin{document}
\title{Electronic structure of half-metallic double perovskites 
}

\author{Z. Szotek$^{1}$, W.M. Temmerman$^{1}$, A. Svane$^{2}$, L. Petit$^{2}$, and H. Winter$^{3}$}
\affiliation{$^{1}$ Daresbury Laboratory, Daresbury, Warrington, WA4 4AD, Cheshire, U.K. \\
$^{2}$ Institute of Physics and Astronomy, University of Aarhus, DK-8000 Aarhus C, Denmark\\
$^{3}$ INFP, Forschungszentrum Karlsruhe GmbH, Postfach 3640, D-76021 Karlsruhe, Germany}

\date{\today}

\begin{abstract}
We present the self-interaction corrected local spin density (SIC-LSD) electronic 
structure and total energy calculations, leading also to valencies of the ground state 
configurations, for the half-metallic double perovskites such as Sr$_{2}$FeMoO$_{6}$, 
Ba$_{2}$FeMoO$_{6}$, Ca$_{2}$FeMoO$_{6}$, and Ca$_{2}$FeReO$_{6}$. We conclude that 
the Fe and Mo (or Re) spin magnetic moments are anti-parallel aligned, and the magnitude 
of the hybridization induced moment on Mo does not vary much between the different compounds. 
The hybridization spin magnetic moment on Re is of the order of -1.1 $\mu_{B}$, while that on
Mo is about -0.4 $\mu_{B}$, independently of the alkaline earth element. 
Also the electronic structure of all the compounds studied is very similar, with a well 
defined gap in the majority spin component and metallic density of states for the 
minority spin component, with highly hybridized Fe, Mo (or Re), and oxygen bands.
\end{abstract}


\maketitle

\section{Introduction}
The interest in double perovskites A$_{2}$B'B''O$_{6}$, with A being alkaline
earth (AE) ion, and the transition metal sites (B-sites) alternately occupied
by different cations B' and B'', has really taken off with the paper by Kobayashi et 
al. \cite{kobayashi}, demonstrating a convincing half-metallic behaviour in 
Sr$_{2}$FeMoO$_{6}$. The calculated electronic structure of this compound shows a
gap in the majority spin component, while for the minority spin component one observes
strongly hybridized Fe 3$d(t_{2g})$, Mo 4$d(t_{2g})$, and O $2p$ states at the Fermi
level. Their study has also established that the latter compound exhibits
tunnelling magnetoresistance (TMR) at low applied magnetic field even at room 
temperature. The double perovskites were discovered in 1960s 
\cite{longo,sleight,patter,galasso,sleight1}
and seem to be metallic and ferromagnetic for B'=Fe, and B''=Mo or Re. 
Unlike the manganites, they are stoichiometric and have large Curie temperatures, T$_{c}$, 
which makes them very promising for technological applications. The Sr$_{2}$FeMoO$_{6}$ 
compound has T$_{c}$ of about 450 K, but also Ba$_{2}$FeMoO$_{6}$ and Ca$_{2}$FeMoO$_{6}$ 
have been shown to have interesting TMR characteristics and high T$_{c}$'s.\cite{coey} 
Although Sr$_{2}$FeReO$_{6}$ (see Ref. \onlinecite{SrRe}) shows similar TMR properties 
to those of Sr$_{2}$FeMoO$_{6}$, recent studies claim that the Ca-compound, Ca$_{2}$FeReO$_{6}$, 
unlike Ca$_{2}$FeMoO$_{6}$, may be insulating.\cite{Millis} It has been suggested that the 
magnetic temperature T$_{c}$, and whether the ground state is metallic or insulating, vary as
A is changed from Ba to Sr to Ca. \cite{Millis1} Also mis-site disorder may have pronounced
effect on the magnetic properties of these compounds. \cite{ogale,Tanusri}

Recent experiments\cite{besse} have established in Sr$_{2}$FeMoO$_{6}$ an electronic structure
and configuration of five localized majority electrons (high spin state) on Fe and
one delocalized electron shared between Mo and the other sites. We aim to validate
this electronic structure by a more systematic study involving two other alkaline earths,
Ca and Ba, as well as Sr. The aspect of a delocalized Mo electron is investigated by replacing
Mo by Re and investigating the amount of delocalized electrons shared between Re and the other sites.
 
In this paper we present an application of the self-interaction corrected local
spin density (SIC-LSD) approximation to the double perovskites Ba$_{2}$FeMoO$_{6}$,
Ca$_{2}$FeMoO$_{6}$, Sr$_{2}$FeMoO$_{6}$, and Ca$_{2}$FeReO$_{6}$. Specifically,
we concentrate on the electronic and magnetic properties of these compounds and in 
particular the size and relative orientation of the spin moments of Fe and Mo (or Re). 
In all the compounds the SIC-LSD calculations find a spin moment of about -0.4 $\mu_{B}$ 
on the Mo and about -1.1 $\mu_{B}$ on Re, with an opposite orientation to the respective 
spin moments on Fe sites. 
In addition, we find these double perovskites to be half-metallic, with a well defined gap 
in the spin-up density of states, and strong hybridization at the Fermi energy between the 
spin-down Fe 3$d$, Mo 4$d$ (or Re 5$d$), and O 2$p$ states. 

The earlier first-principles calculations performed for double perovskites had to be done
with GGA and implement theoretical optimisation of oxygen positions to obtain the 
half-metallic groundstate. \cite{kobayashi,sarma} Other applications were done with the
LDA+U, \cite{fang} which similarly to SIC-LSD provides better description of localized
states in solids. The present study is the first SIC-LSD application to double perovskites.
Its advantage is that it treats both localized and delocalized states on equal footing,
and therefore should be well suited for describing the localized nature of Fe $3d$ electrons
as well as the Fe valency.

The remainder of the paper is organized as follows. In the next section, we briefly present the 
theoretical background of the SIC-LSD approach. In section III, the technical and computational
details regarding the application to double perovskites are elaborated upon. In section IV, we 
present total energy and density of states calculations. There we also compare magnetic moments 
and determine valency of Fe in the studied compounds. The paper is concluded in section V.

\section{Theory}

The basis of the SIC-LSD formalism is a self-interaction free total energy functional, 
\( E^{SIC} \), obtained by subtracting from the LSD total energy functional,
\( E^{LSD} \), a spurious self-interaction of each occupied electron state 
\( \psi _{\alpha } \)\cite{PZ81}, namely 
\begin{equation}
\label{eq1}
E^{SIC}=E^{LSD}-\sum _{\alpha }^{occ.}\delta _{\alpha }^{SIC}.
\end{equation}
 Here \( \alpha  \) numbers the occupied states and the self-interaction correction
for the state \( \alpha  \) is 
\begin{equation}
\label{eq2}
\delta _{\alpha }^{SIC}=U[n_{\alpha }]+E_{xc}^{LSD}[\bar{n}_{\alpha }],
\end{equation}
with \( U[n_{\alpha }] \) being the Hartree energy and \( E_{xc}^{LSD}[\bar{n}_{\alpha }] \)
the LSD exchange-correlation energy for the corresponding charge density \( n_{\alpha } \)
and spin density \( \bar{n}_{\alpha } \). It is the LSD approximation to the exact 
exchange-correlation energy functional which gives rise to the spurious self-interaction. 
The exact exchange-correlation
energy \( E_{xc} \) has the property that for any single electron spin density,
\( \bar{n}_{\alpha } \), it cancels exactly the Hartree energy, namely \begin{eqnarray}
U[{n}_{\alpha }]+E_{xc}[\bar{n}_{\alpha }]=0.\label{can}
\end{eqnarray}
In the LSD approximation this cancellation does not take place, and for truly localized
states the correction  (Eq. (\ref{eq2})) can be substantially different from zero. For extended states in
periodic solids the self-interaction vanishes.
Consequently, the SIC-LSD approach can be viewed as a genuine extension of LSD
in the sense that the self-interaction correction is only finite for spatially
localized states, while for Bloch-like single-particle states \( E^{SIC} \)
is equal to \( E^{LSD} \). Thus, the LSD minimum is also a local minimum of
\( E^{SIC} \). 

It follows from minimization of Eq. (\ref{eq1}) that within the SIC-LSD approach
such localized electrons move in a different potential than the delocalized
valence electrons which respond to the effective LSD potential. For example,
in the case of the double perovskites, five (Fe$^{3+}$) or six (Fe$^{2+}$)
Fe $d$ electrons move in the SIC potential, while all other electrons feel 
only the effective LSD potential. Thus, by including
an explicit energy contribution for an electron to localize, the ab-initio SIC-LSD
describes both localized and delocalized electrons on an equal footing, leading
to a greatly improved description of static Coulomb correlation effects over
the LSD approximation. Assuming various atomic configurations, consisting of
different numbers of localized states, one can explore the corresponding local
minima of the SIC-LSD energy functional of Eq. (\ref{eq1}), and determine the
lowest energy solution and valency. The advantage of the SIC-LSD formalism is that 
for such systems as transition metal oxides or rare earth compounds the lowest 
energy solution will describe the situation where some single-electron states 
may not be of Bloch-like form. Regarding the valency, it is determined here as 
the integer number of electrons available for band formation, i.e.,
\[
N_{val}=Z-N_{core}-N_{SIC},
\]
where $Z$ is the atomic number (26 for Fe), $N_{core}$ is the number of core (and semi-core)
electrons (18 for Fe), and $N_{SIC}$ is the number of localized, i.e., self-interaction
corrected, states (either five or six, respectively for Fe$^{3+}$ and Fe$^{2+}$).

In the present work the SIC-LSD approach has been implemented\cite{TSSW98} 
within the linear muffin-tin-orbital (LMTO) band structure method\cite{oka75} in the 
tight-binding representation\cite{AJ84}, where the electron wave functions are expanded in
terms of the screened muffin-tin orbitals, and the minimization of \( E^{SIC} \)
becomes a non-linear problem in the expansion coefficients.  The atomic spheres
approximation (ASA) has been used, according to which the polyhedral Wigner
Seitz cell is approximated by slightly overlapping atom centered spheres, with a
total volume equal to the actual crystal volume.

\section{Calculational details}

All the compounds studied here crystallize in ordered double perovskite structures. The 
simplest system is Ba$_{2}$FeMoO$_{6}$ which occurs in the cubic structure, with the 
Fm3m space group. The Sr$_{2}$FeMoO$_{6}$ compound has the body-centered tetragonal 
{\it bct} structure, with the I4/mmm space group. Replacing Sr by a smaller still Ca 
element results in a monoclinic structure for the Ca$_{2}$FeMoO$_{6}$ and 
Ca$_{2}$FeReO$_{6}$ compounds, with the P2$_{1}$/n space group. The experimental lattice 
parameters used in the calculations for the first three compounds have been taken from
Refs. \onlinecite{coey} and \onlinecite{kobayashi}, and for the last compound from 
Ref. \onlinecite{Millis}. For the Ca compounds two formula units (20 atoms) have been
necessary. In addition, for a better space filling also eight empty spheres have been 
introduced in Re-compound. 
Concerning the linear muffin-tin basis 
functions, we have used $4s$, $4p$, and $3d$ partial waves on all Fe atoms, and treated 
them as low-waves, on the oxygens ($2s$ and $2p$) and Ca ($4s$ and $3p$) in addition to
the $s$ and $p$ low partial waves, $3d-$partial waves have been treated as intermediate. 
On the empty spheres the $1s-$waves have been considered as low- and the $2p$'s as 
intermediate-waves. On the Ba ($6s$, $5p$, and $5d$), Sr ($5s$, $4p$, and $4d$), 
Mo ($5s$, $5p$, and $4d$) and Re ($6s$, $6p$, and $5d$) atoms, the $s-$, $p-$, and $d-$waves
have been treated as low and the $4f-$waves as intermediate. In all the cases, the 
SIC-LSD calculations have been started using the self-consistent LSD charge densities 
with the Perdew and Zunger exchange-correlation potential. \cite{PZ81} Mostly the Fe 3$d$ 
electrons have been treated as localized states (in some test calculations we have also
treated Mo $d$'s as localized), meaning that they have been moving in 
the self-interaction corrected LSD potentials, while all other electrons have seen the 
effective LSD potential. In all the cases studied, we have allowed for a number of 
different configurations of SIC 3$d$ states to find the ground state solution. Namely, 
treating five 3$d$ electrons on the Fe's as localized, moving in the SIC-LSD effective 
potentials, meant that the Fe ions were trivalent. Similarly, for divalent Fe's, six 
3$d$ electrons on the Fe's were considered to be localized. By comparing total energies 
corresponding to different localized Fe 3$d$ configurations we have been able to find the 
ground state configuration which for all the compounds has been that of Fe$^{3+}$. 
However, other configurations have been energetically close, with the Fe$^{2+}$ being 
insulating in all Mo-based compounds. Although, unlike in the work by Kobayashi et al.,
\cite{kobayashi} optimizing oxygen positions in Sr$_{2}$FeMoO$_{6}$ was not 
needed for obtaining half-metallic ground state, however in order to study its effect 
on the magnetic properties and to make contact with Ref. \onlinecite{kobayashi}, for 
this compound we have also performed calculations for the scenario where the theoretically 
optimized positions of Ref. \onlinecite{kobayashi} have been implemented. \cite{besse}

\section{Results and discussion}


The SIC-LSD spin-polarized total densities of states (DOS) per formula unit are presented 
in Fig. 1 for Ca$_{2}$FeReO$_{6}$, Ca$_{2}$FeMoO$_{6}$, Sr$_{2}$FeMoO$_{6}$, and
Ba$_{2}$FeMoO$_{6}$ compounds. There we also show the O $2p$ spin polarized densities of
states. The spin polarized densities of states of Fe $3d$ and Mo $4d$ and Re $5d$ are
given in Fig. 2, for all the different compounds. First thing to note is 
the similarity between the densities of states, 
independently of which alkaline earth compound is
considered. For all compounds we see convincing half-metallic behaviour, with well 
defined gap at the Fermi energy in the majority spin channel, and strongly hybridized 
Fe $3d$, Mo $4d$ (or Re $5d$) and oxygen $2p$ states, in the other spin channel. 
The half-metallic character is reflected in the total spin magnetic moments that are
all integer, as can be seen in Table I, where we summarize the total and species
decomposed spin moments for all the compounds together with the respective volumes
per formula unit. As can be seen in Fig. 1, the valence band consists predominantly 
of O $2p$ states, but contains a small admixture of Mo, Re- and Fe $d-$states.
The unoccupied majority Re $d$ bands lie closer to the Fermi energy in comparison
with the unoccupied majority Mo $d$ states in Ca$_{2}$FeMoO$_{6}$. We can also
see that the unoccupied Mo $d$ states move closer to the Fermi energy as one
traverses the series in the sequence Ca, Sr and Ba. However, the majority
spin channel maintains a band gap for all these compounds. In the minority
spin channel these systems are metallic with O 2p, Mo 4d and Fe 3d states
straddling the Fermi energy. These are the states which reduce the Fe magnetic moment
from essentially 5 $\mu_{B}$ to 3.87 $\mu_{B}$ (Ca$_{2}$FeReO$_{6}$), 3.76 $\mu_{B}$
(Ca$_{2}$FeMoO$_{6}$), 3.71 $\mu_{B}$ (Sr$_{2}$FeMoO$_{6}$) and 3.81 $\mu_{B}$
(Ba$_{2}$FeMoO$_{6}$) (see Table I). From these DOS one can also expect the oxygen
to acquire a moment. This happens according to Table I and it is parallel to the
Fe spin moment and is non-neglegible at around 0.1 $\mu_{B}$. 
However it is the Mo and Re sites which are of interest.
Substantial spin magnetic moments of -0.4 $\mu_{B}$ and -1.1 $\mu_{B}$
are found on the Mo and Re sites respectively. This results from the strong
hybridization with Fe in the minority spin channel. This also explains 
the antiparallel alignment of these spin magnetic moments with respect
to Fe.

\begin{table}
\caption{Total and species decomposed spin magnetic moments (in $\mu_{B}$)
as calculated self-consistently within SIC-LSD band structure method. In
the last row the respective volumes per formula unit are quoted (in 
(atomic unit)$^{3}$). }
\begin{tabular}{ccccc}
Moment & Ca$_{2}$FeReO$_{6}$ & Ca$_{2}$FeMoO$_{6}$ & Sr$_{2}$FeMoO$_{6}$ & Ba$_{2}$FeMoO$_{6}$ \\
\hline
M$_{total}$    &  3.00  & 4.00  &  4.00 (4.00) &  4.00 \\
M$_{Fe}$       &  3.87  &  3.76 &  3.71  (3.65) &  3.81 \\
M$_{Mo}$       &   -    & -0.40 & -0.43 (-0.35) & -0.41 \\
M$_{Re}$       & -1.12  &   -   &    -          &    -  \\
M$_{AE}$       & -0.02  &  0.01 &  0.02 (0.02)  &  0.02 \\
M$_{O1}$       &  0.02  &  0.10 &  0.11 (0.11)  &  0.09 \\
M$_{O2}$       &  0.01  &  0.11 &  0.11 (0.11)  &    -  \\
M$_{O3}$       &  0.11  &  0.11 &    -          &    -  \\
Volume         & 773.8  & 777.5 & 830.2         & 884.0 \\
\end{tabular}
\label{table1}
\end{table}

The densities of states in the majority spin channels for these four compounds
are insulating, since a fully occupied O $p$ band and five localized Fe $d$ states 
accomodate 23 (6$\times$3+5=23) valence electrons. Of the remaining 
19 or 20 valence electrons for the Mo and Re compounds respectively, 18
fill up the minority O $p$ band. This leaves one or two electrons 
for the Mo and Re compounds to occupy the minority states which straddle the
Fermi level, and have Fe $d$, Mo or Re $d$ and oxygen $p$ character. We have also 
considered a scenario of localising through the self-interaction
correction one Fe (Mo) $d$ state in the minority spin channel describing Fe$^2+$
(Mo$^5+$) respectively. In this case the states straddling the Fermi energy which
accomodate this one electron are pulled down below the bottom of the O $p$ valence
bands and we also obtain an insulating state in the minority spin channel. 
However, these scenarios are energetically unfavourable by 70 mRy and 103 mRy 
for Mo$^{5+}$ and Fe$^{2+}$ respectively. Obviously the gain in localisation
energy does not compensate the loss in hybridization energy.



It is interesting to see how little the magnetic properties of these compounds are
affected by the size of the alkaline earth atom. There is hardly any variation in the
size of the Fe and Mo or Re moments with the substantial change in volume of the
compounds (see Table I). Also the reduction in the size of the Fe and Mo spin moments for 
Sr$_{2}$FeMoO$_{6}$, as a result of implementing theoretical optimization of oxygen 
positions, as given by Kobayashi et al.,\cite{kobayashi} does not seem too important 
although brings the size of our calculated moments close to the values of 
Ref. \onlinecite{kobayashi}. The reduced 
values are given in parenthesis in the fourth column of Table I. Concerning spin moments
of the other species, the small changes cannot be resolved when rounded off to two 
digits after the decimal point.

Finally, we would like to comment on the valency of Fe and the ground state configuration
for all the compounds. Regarding energetics, in all SIC-LSD calculations the Fe$^{3+}$
configuration has been most energetically favourable, followed by  Fe$^{4+}$ (energetically
unfavourable by 53, 61 and 83 mRy in Ba-, Sr- and Ca-compounds, respectively), and  Fe$^{2+}$
solution, at least twice as unfavourable as the former configuration (specifically, by 103, 
122 and 250 mRy). Moreover, treating also Mo $d$ states as localized has been energetically
unfavourable, and the more states have been localized the more unfavourable solutions
have been obtained.
\section{Conclusions}

In summary, the SIC-LSD approach has been applied to four different double perovskites,
and in all cases the half-metallic ground state solutions have been obtained. 
This seems to be a rather generic result for double perovskites involving Fe
with a filled majority d shell and transition metal ions which donate few, one (Mo) 
to two (Re), d electrons. This type of materials could therefore be a fruitful playground
for the discovery of new spintronics materials.
In particular,
also the Ca$_{2}$FeReO$_{6}$ has been found to be half-metallic, and not insulating as
indicated in Ref. \onlinecite{Millis}. The electronic structure for all studied compounds
appears to be very similar, with a gap in the majority spin component, and strongly 
hybridized bands at the Fermi level for the minority spin component. The spin magnetic
moments induced by hybridization on the Mo sites have been of the similar magnitude, 
about -0.4 $\mu_{B}$, independently of the alkaline earth element, while that of Re is
of the order of -1.1 $\mu_{B}$. These induced spin magnetic moments are antiparallel 
aligned with the Fe spin moment. Concerning the Fe valency, for all the compounds the
trivalent configuration has been most favourable, followed by the tetravalent and
divalent ones. 

\section*{Acknowledgements}
We thank Dr. Tanusri Saha-Dasgupta for useful communications and Professor P.H.
Dederichs for valuable discussions.\\


\begin{figure*}
\begin{tabular}{cc}
\includegraphics[scale=.37,angle=-90]{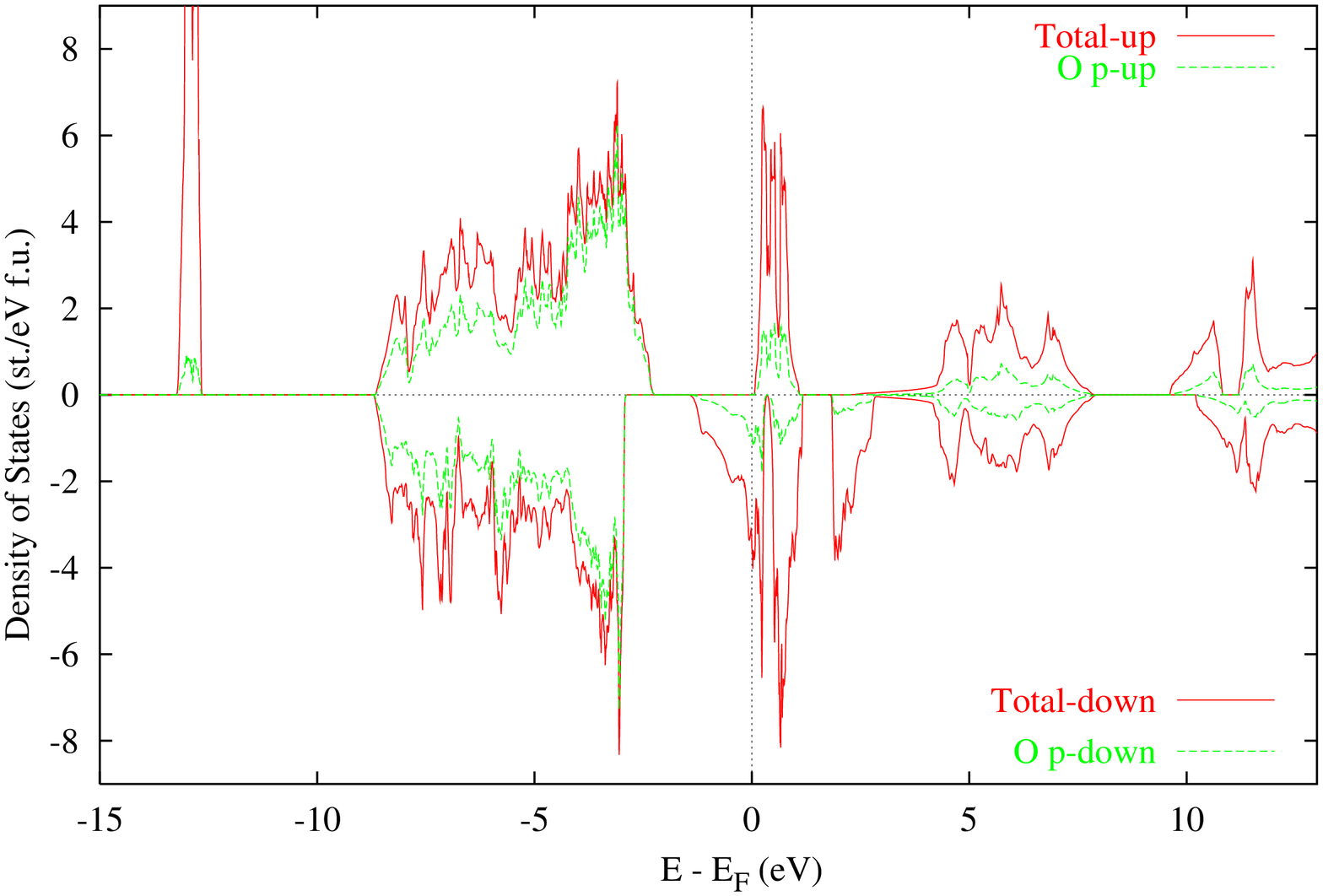}&
\includegraphics[scale=.37,angle=-90]{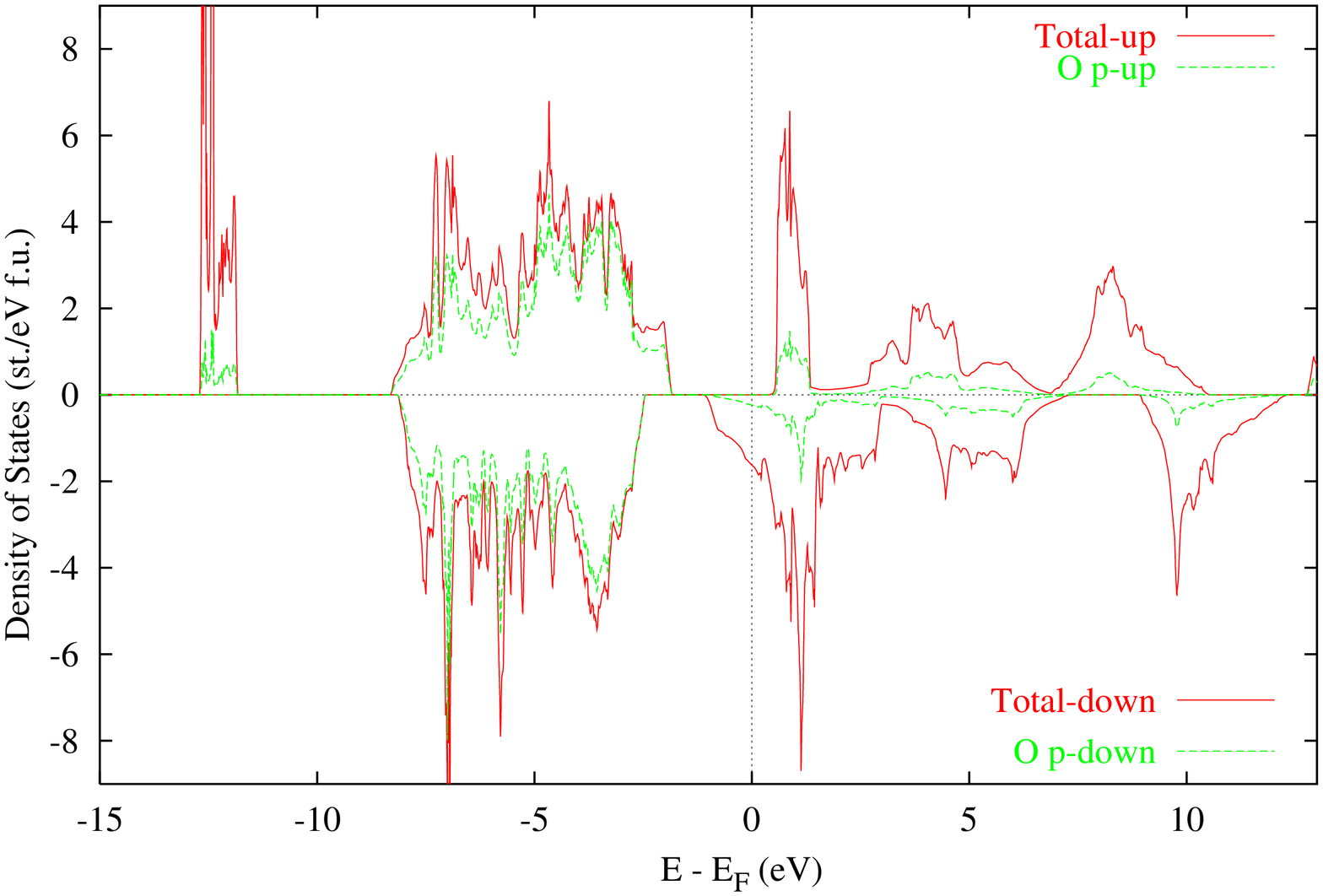}\\
\includegraphics[scale=.37,angle=-90]{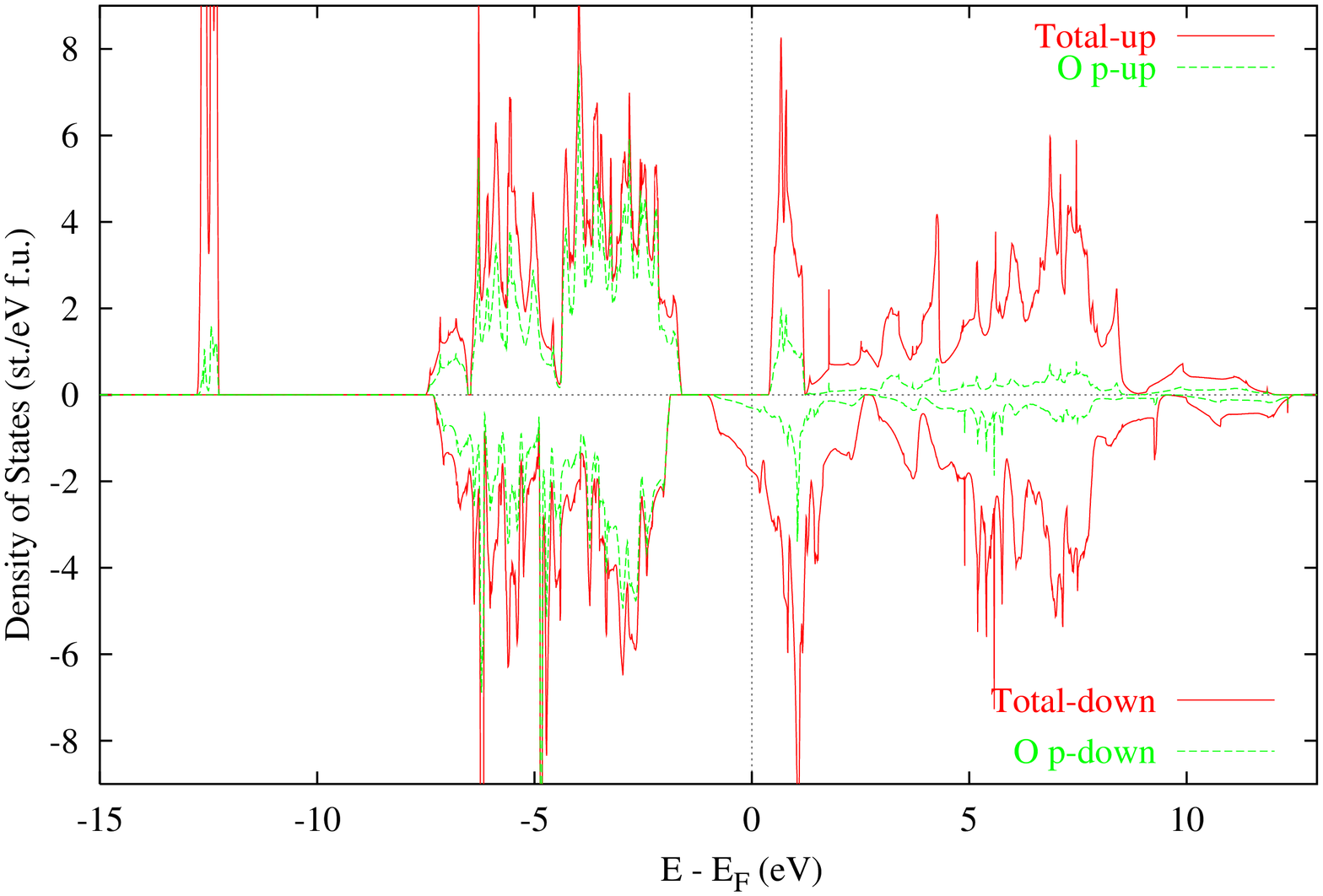}&
\includegraphics[scale=.37,angle=-90]{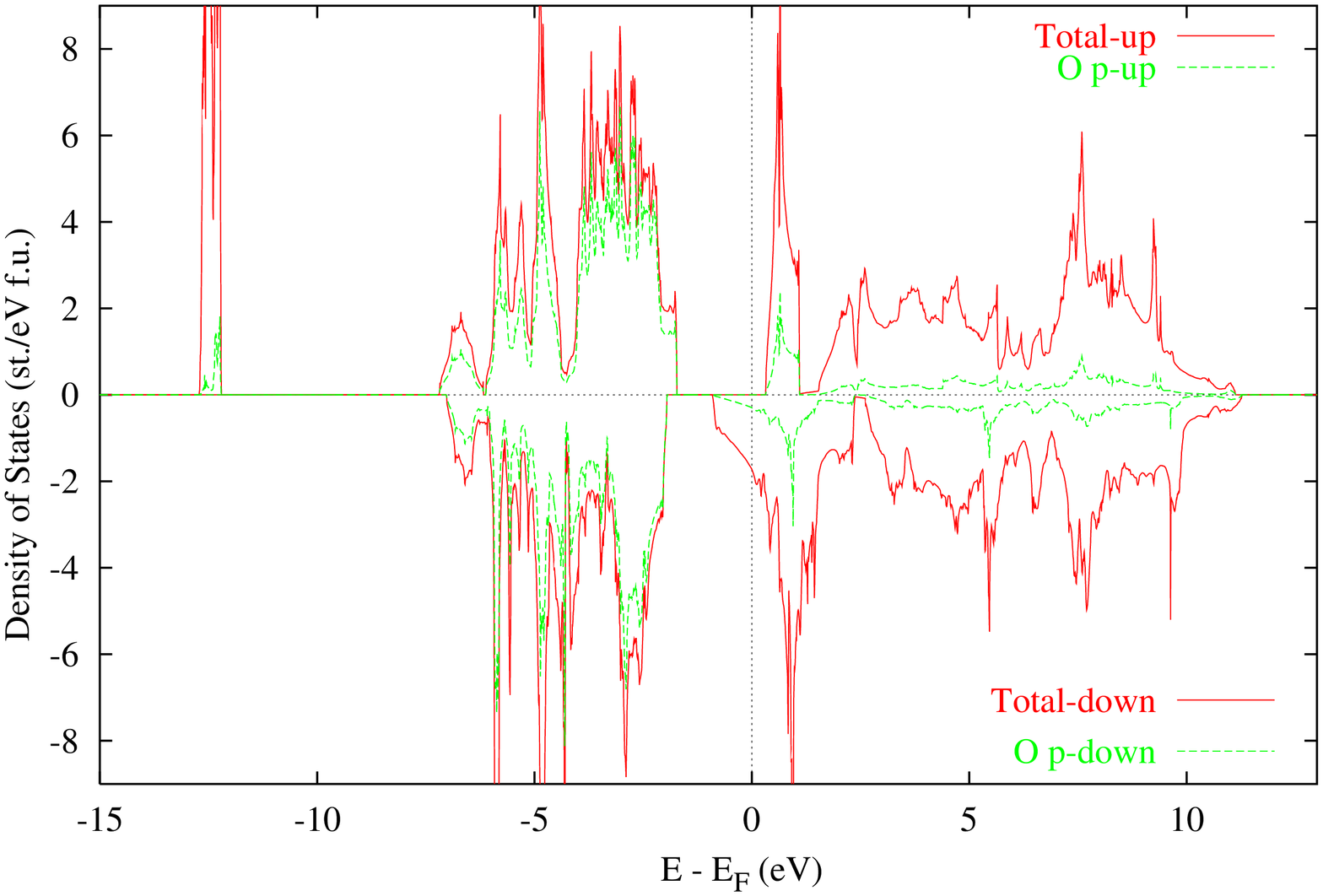}
\end{tabular}
\caption{Spin polarized total and oxygen $2p$ densities of states
per formula unit, for Ca$_{2}$FeReO$_{6}$ (top row, left), Ca$_{2}$FeMoO$_{6}$ (top row, right),
Sr$_{2}$FeMoO$_{6}$ (bottom row, left), and Ba$_{2}$FeMoO$_{6}$ (bottom row, right).}
\label{Fig1}
\end{figure*}

\begin{figure*}
\begin{tabular}{cc}
\includegraphics[scale=.37,angle=-90]{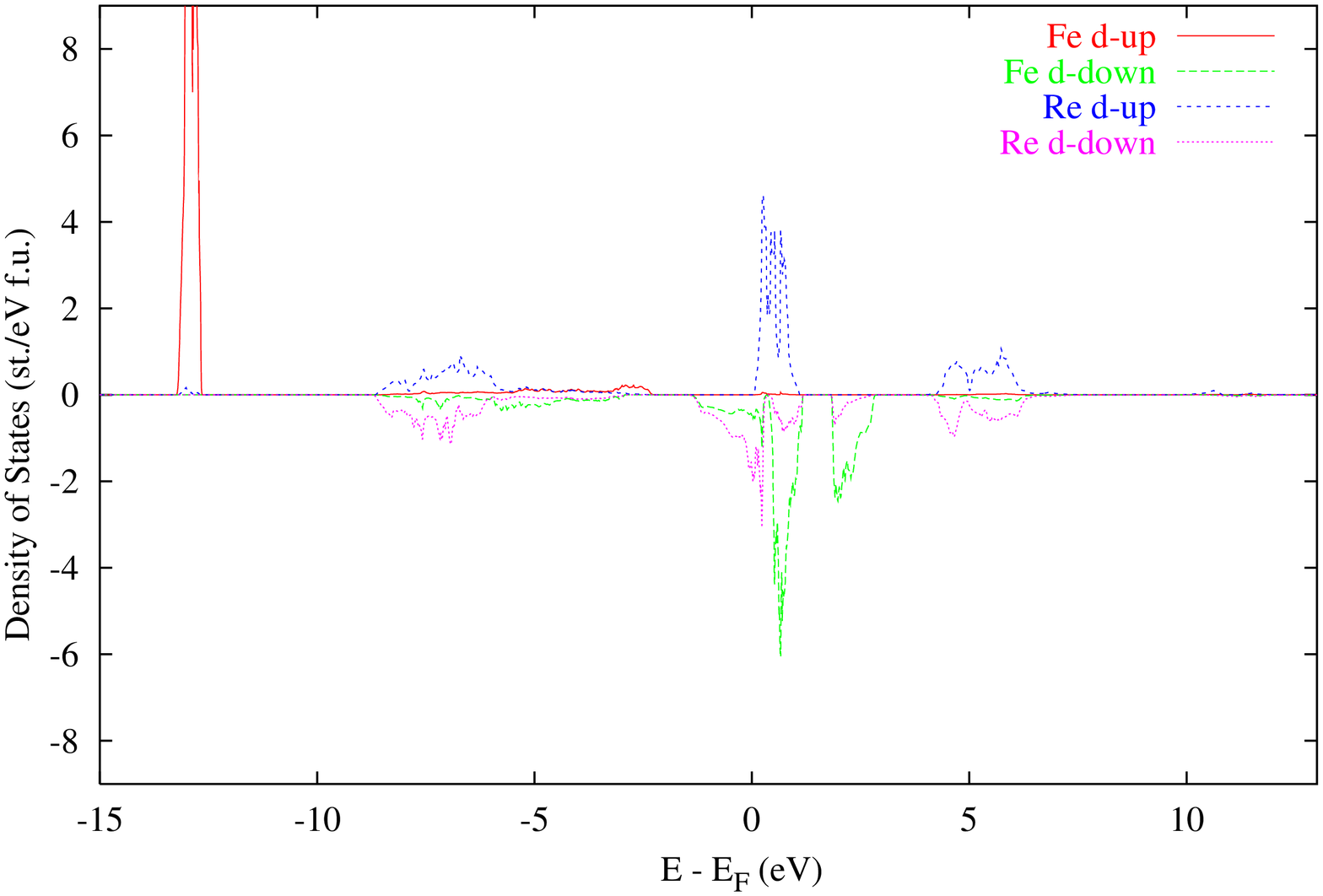}&
\includegraphics[scale=.37,angle=-90]{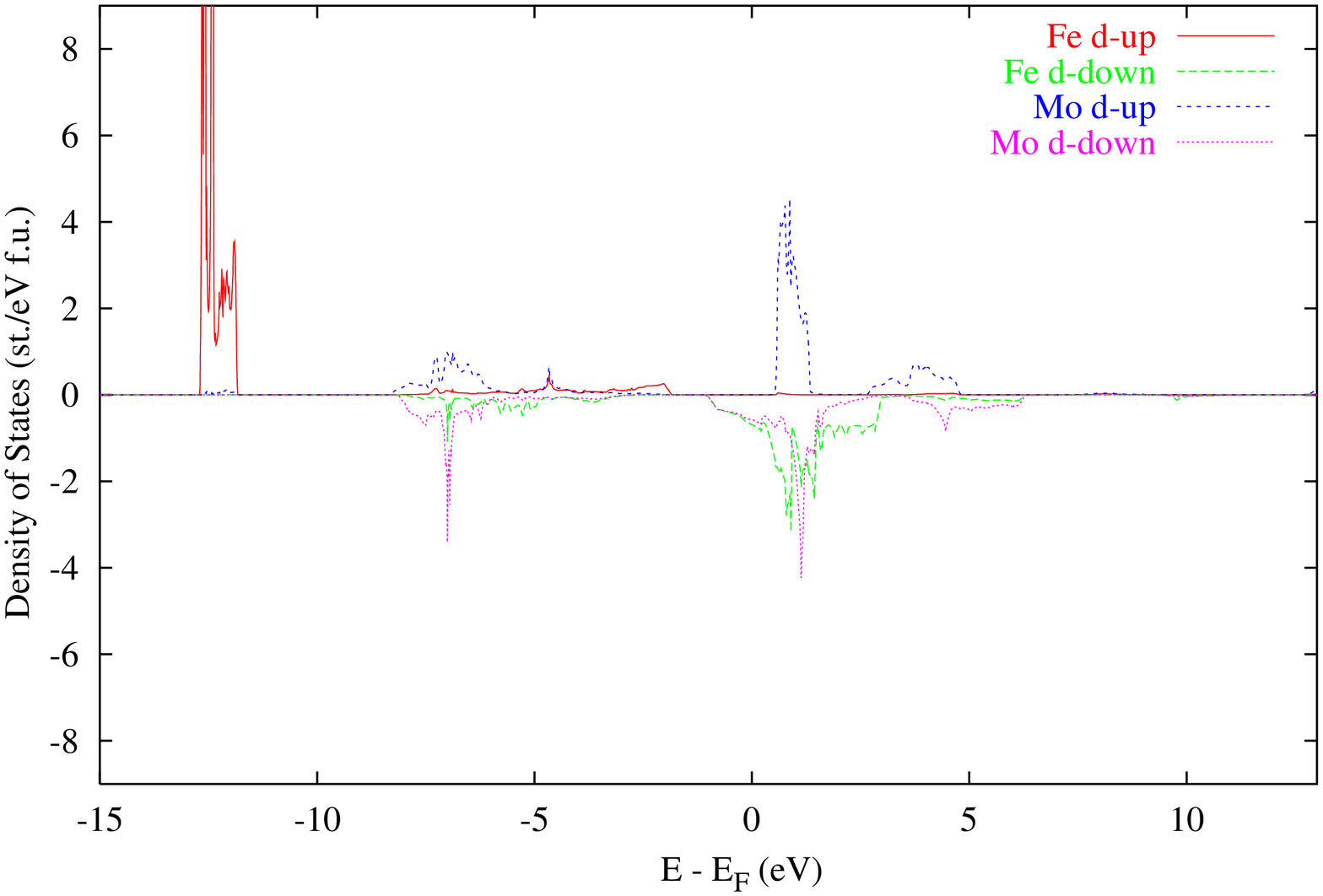}\\
\includegraphics[scale=.37,angle=-90]{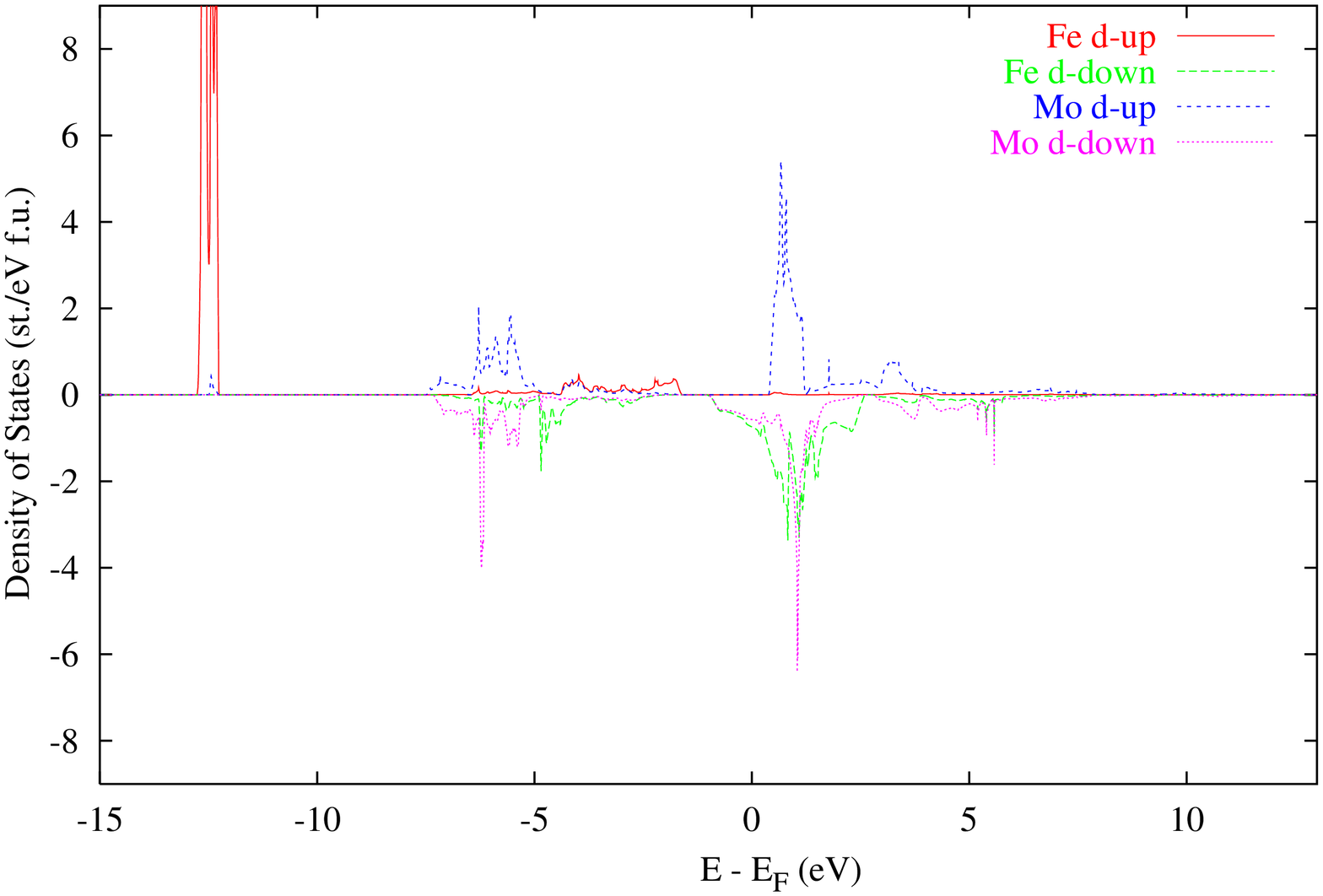}&
\includegraphics[scale=.37,angle=-90]{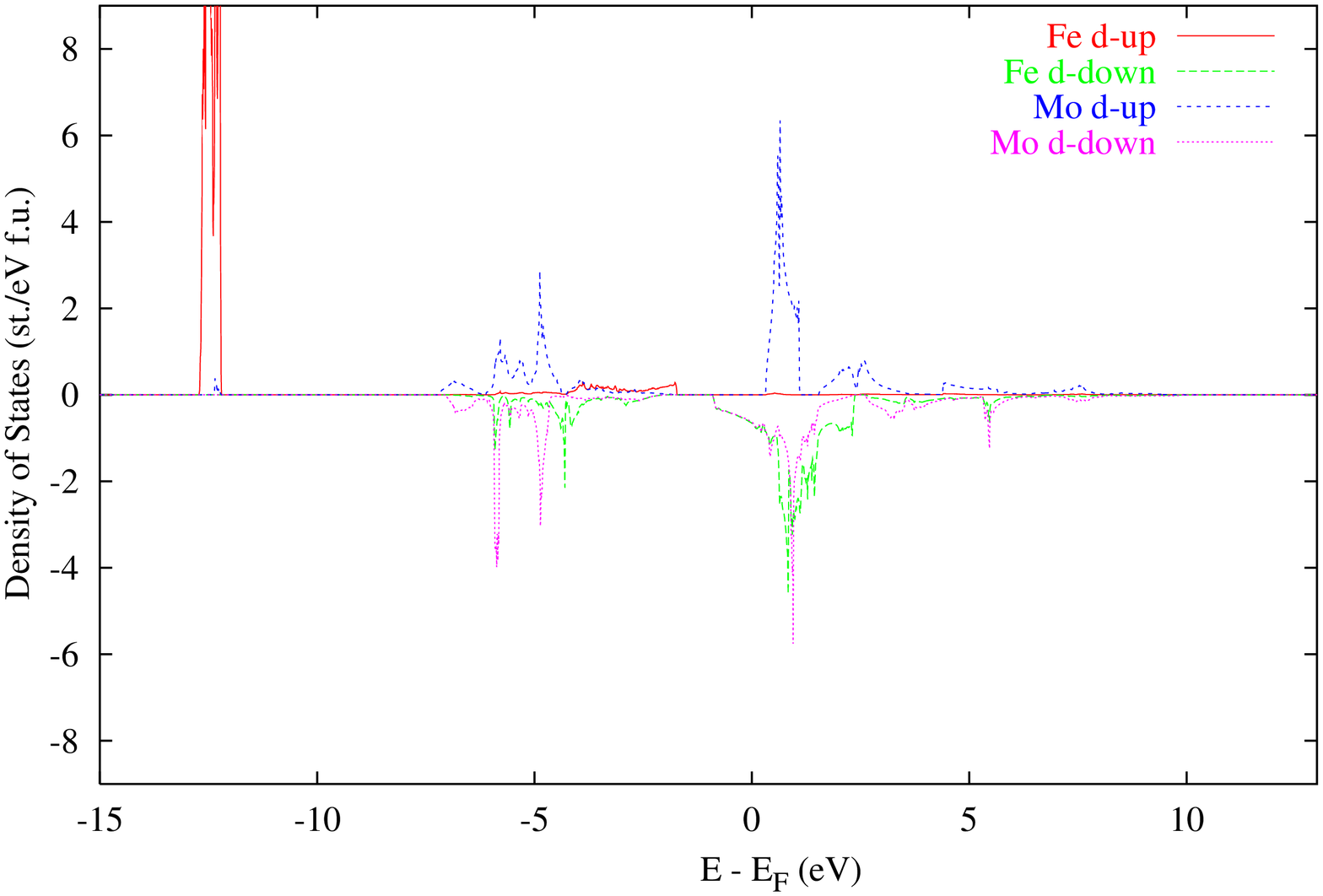}
\end{tabular}
\caption{Spin polarized Fe $3d$ and Re $5d$ densities of states per formula
unit in Ca$_{2}$FeReO$_{6}$ (top row, left), and spin polarized Fe $3d$ and Mo $4d$ 
densities of states per formula unit in Ca$_{2}$FeMoO$_{6}$ (top row, right),
Sr$_{2}$FeMoO$_{6}$ (bottom row, left), and Ba$_{2}$FeMoO$_{6}$ (bottom row, right).}
\label{Fig2}
\end{figure*}


\end{document}